\documentclass[12pt]{article}

\usepackage{amssymb}
\usepackage{amsmath}
\usepackage{graphicx}
\usepackage{latexsym}

\newcommand{\gsim}{ \mathop{}_{\textstyle \sim}^{\textstyle >} }

\begin{document}

\setcounter{footnote}{0}
\begin{titlepage}

\begin{center}


\vskip .5in

{\Large \bf
Stability of Metastable Vacua 
in Gauge Mediated SUSY Breaking Models 
with Ultra Light Gravitino}

\vskip .45in

{\large
Junji Hisano, Minoru Nagai, Masato Senami and Shohei Sugiyama
}

\vskip .45in

{\em
Institute for Cosmic Ray Research, \\
University of Tokyo, 
Kashiwa 277-8582, Japan}

\end{center}

\vskip .4in

\begin{abstract}
  Recently Murayama and Nomura proposed a simple scheme to construct
  the gauge mediation models, using metastable supersymmetry breaking
  vacua.  It has a possibility to predict the ultra light gravitino
  mass $m_{3/2}\lesssim 16~{\rm eV}$, while such a light gravitino may
  destabilize the metastable vacua. We investigate stability of the
  metastable vacuum of their model.  The transition rate from the
  false vacuum to true ones is evaluated by numerical calculation,
  including the Coleman-Weinberg potential destabilizing the
  metastable vacuum. It is found that when the messenger sector is
  minimal, stability of the metastable vacuum imposes an upperbound on
  squark mass $M_{\tilde q}$ for the ultra light gravitino as
  $M_{\tilde q} \lesssim 1800$~GeV at most.  Squarks
  with this mass may be found in the LHC experiments.
\end{abstract}

\end{titlepage}

\renewcommand{\thepage}{\arabic{page}}
\setcounter{page}{1}

\setcounter{equation}{0}
\section{Introduction}

Introduction of Supersymmetry (SUSY) is one of the most attractive
directions to construct models beyond the Standard Model (SM), since
SUSY cures the hierarchy problem.  However, when SUSY
breaking mass terms of squarks and sleptons are generic in the SUSY
SM, too large flavor changing neutral current (FCNC) processes are
predicted. This problem is called as the SUSY flavor problem.  In
addition, the SUSY SM also has the SUSY CP problem. The SUSY breaking
trilinear and bilinear scalar coupling terms ($A$ and $B$ terms) and
the gaugino mass terms have CP phases generically.  The predictions
for electric dipole moments are rather large, though the constraints
by experiments are severe.

The Gauge Mediated SUSY Breaking (GMSB) models
\cite{Dine:1981za,Dine:1993yw} are an excellent solution to these
problems.  SUSY breaking in a hidden sector is transmitted to the
visible sector through the SM gauge interactions, which are
flavor-blind.  The SUSY breaking mass terms of squarks and sleptons
are then flavor-blind, so that dangerous FCNC processes are naturally
suppressed.  In addition, the phases of the $A$ and $B$ terms are
automatically aligned to those of the gaugino mass terms in the
Minimal Gauge Mediation (MGM) model \cite{Dine:1996xk}, since the $A$
and $B$ terms are vanishing at tree level in the model. In the case,
the SUSY CP problem is avoided\footnote{
  It is shown in Ref.~\cite{Hisano:2007ah} that the latest
  experimental values of the anomalous magnetic moment of the muon and
  the branching ratio of $\bar{B}\to X_s\gamma$ favor a low messenger
  scale ($\sim 100$~TeV) in the MGM model. This is quite consistent
  with the ultra light gravitino discussed in this paper. }.

In the GMSB models, the lightest SUSY particle is the gravitino, with
mass $m_{3/2} \simeq $(1~eV - 10~GeV). An ultra light gravitino
$m_{3/2} \lesssim 16$~eV is known to have no cosmological gravitino
problem, such as overclosure \cite{Pagels:1981ke} and too large
free-streaming lengths \cite{Viel:2005qj}.  However, only a few models
were previously proposed in order to realize the ultra light gravitino
\cite{VLGmodels}.

Recently Murayama and Nomura proposed in Ref.~\cite{SimpGMSB} a simple
scheme to construct the GMSB models along a metastable SUSY breaking
scenario.  Metastable SUSY breaking is recognized as an interesting
alternative SUSY breaking scenario, after Intriligator, Seiberg and
Shih found that even simple vector-like theories, such as SUSY SU(N)
QCD, can break SUSY on metastable local minima
\cite{Intriligator:2006dd}.  Many phenomenological models with
metastable SUSY breaking vacua are nowadays discussed.

The scheme proposed by Murayama and Nomura has a quite simple
structure and a possibility of the ultra light gravitino mass
$m_{3/2}\lesssim 16~{\rm eV}$.  However, the stability of the
metastable vacua becomes weaker for smaller gravitino
mass.  It is not clear whether the lifetimes of the metastable vacua are
longer than the age of the universe.

In this paper, we study stability of the SUSY breaking metastable
vacuum in the GMSB model which is realized in their scheme as an effective
theory \cite{SimpGMSB},  in the case of the ultra light gravitino
mass. The lifetime of the metastable vacuum is numerically evaluated
in the semiclassical techniques \cite{Coleman}. In the evaluation we
include the Coleman-Weinberg (CW) potential in the scalar potential,
which destabilizes the metastable vacuum.  In Ref.~\cite{SimpGMSB},
the authors estimated the lifetime by a triangular approximation
\cite{Duncan:1992ai}. However, this approximation is not accurate
especially in this model, since several scalar fields are relevant to
the phase transition, as the authors mentioned in their text.  In the
case, numerical calculation is required to estimate a reliable value
of the transition rate.

We also derive upperbounds on the squark mass and the cutoff scale for
ultraviolet (UV) models for $m_{3/2}\lesssim 16~{\rm eV}$.  The bound
on the gravitino mass can be translated to those on the squark and
gluino masses in this model, since the visible sector is more directly
coupled with the hidden sector in this model.

This paper is organized as follows.  In the next section, we review
the model discussed in Ref.~\cite{SimpGMSB}.  In Section 3, we discuss
the stability of the metastable vacuum and constraints on parameters
of this model.  Section 4 is devoted to summary.

\section{Model}

In this section we review the GMSB model with the metastable vacuum
discussed in Ref.~\cite{SimpGMSB}. This model is realized in their scheme to
construct the GMSB model, and has a quite simple structure. The
relation between the gravitino mass and the stability of the
metastable SUSY breaking vacuum is also clarified here.

The superpotential and K$\ddot{\rm a}$hler potential in this model are
given by
\begin{eqnarray} \label{eq:superpotential}
 W = -\mu^2 S + \kappa S f \bar{f} - M f \bar{f},
\end{eqnarray}
\begin{eqnarray} \label{eq:kahler}
 K = |f|^2 + |{\bar f}|^2 + |S|^2 - \frac{|S|^4}{4\Lambda^2} 
           + {\cal O}\left(\frac{|S|^6}{\Lambda^4}\right),
\end{eqnarray}
where $S$ is a gauge singlet chiral field and $f$ and $\bar{f}$ are
messenger fields, which transform as ${\bf 5}$ and ${\bf 5}^*$ under
the SU(5)$_{\rm GUT}$ gauge symmetry, respectively.  Here, $\mu$, $\kappa$ and
$M$ can be taken to be real and positive without loss of generality,
which are taken as free parameters in this model.

This model is an effective theory below the UV cutoff scale $\Lambda$.
It is assumed that higher terms of $|S|$ appear after some particles
with cutoff-scale mass are integrated out.  The negative sign of the
$|S|^4$ term is required for deriving a local minimum near the origin.
The expansion in the above K$\ddot{\rm a}$hler potential implies that
$S$ has some approximate conserved charge.  For example, we can assign
$U(1)_R$ charges of 2, 0 and 0 to $S$, $f$ and $\bar{f}$,
respectively, although the last term of Eq.~(\ref{eq:superpotential})
violates this symmetry explicitly.

The tree-level scalar potential is given from
Eqs.~(\ref{eq:superpotential}) and (\ref{eq:kahler}) as
\begin{eqnarray} \label{eq:potential_tree}
 V_{\rm tree}
  = \left|\mu^2-\kappa f\bar{f}\right|^2 
    \left[ 1+\frac{|S|^2}{\Lambda^2}
             +{\cal O}\left(\frac{|S|^4}{\Lambda^4}\right)\right]
  + \left|\kappa S -M\right|^2( |f|^2 + |\bar{f}|^2)
  + \sum_i \frac12 g_i^2 D_i^2.
\end{eqnarray}
Here, we use the same notation for scalar fields as superfields, and
$D_i$ is the $D$ term for the SM gauge interactions $i$.  Let us 
discuss the vacuum structure in this potential in the following.

First, this potential has global SUSY conserving minima at
\begin{eqnarray} \label{eq:globalmin}
 S = \frac{M}{\kappa},\quad 
 |f| = |\bar{f}| = \frac{\mu}{\kappa^{1/2}},
\end{eqnarray}
where the phases of $f$ and $\bar f$ are constrained as $\arg (f \bar
f ) = 0$.  In this paper, for justifying the expansion of
Eq.~(\ref{eq:kahler}) by $S/\Lambda$, we impose the following
condition,
\begin{eqnarray} \label{eq:constraint_Lambda} \Lambda \gtrsim
  \frac{M}{\kappa} .
\end{eqnarray}
In the case that this condition is not satisfied, the global minimum
(\ref{eq:globalmin}) is not determined within validity of this
potential and dependent on UV models, so that we could not discuss the
stability of the vacuum.

Next, this potential also has a local minimum at the origin of the
field space, $S=f=\bar{f}=0$. SUSY is broken in this vacuum due to
non-zero $F$ term, $F_S = \mu^2$.  The mass squared for the scalar
component of the singlet field $S$ is given by
$m_S^2=\mu^4/\Lambda^2$.  The stability of this metastable vacuum
becomes weaker for larger $\Lambda$, since the potential for $S$
direction becomes flat.

The masses for the fermionic and scalar components of the messenger
fields are given by $M_{\rm mess}=M$ and $m_{f_{\pm}}^2=M^2\pm \kappa
\mu^2$, respectively, around the origin. This minimum is
metastable when the following condition is satisfied,
\begin{eqnarray}
M^2 > \kappa \mu^2 .
\label{eq:constraint_f}
\end{eqnarray}
When $M^2$ is close to $\kappa \mu^2$, the potential for the direction
along $f=\bar{f}$ is flat and this metastable vacuum becomes unstable.

The spontaneous SUSY breaking is link to the $U(1)_R$ symmetry
\cite{Nelson:1993nf}. However, this symmetry is approximate in this
model since the SUSY mass terms of messenger fields violate it. The
radiative correction from messenger fields destabilizes the
SUSY breaking metastable vacuum. Thus, it is essential to include the
radiative correction in discussing the stability of the metastable vacuum.

The messenger fields induce the following CW effective
potential,
\begin{eqnarray} \label{eq:potential_CW}
 V_{\rm CW} 
 &\simeq& 
  \frac{\kappa^2 \mu^4}{16\pi^2}
  F\left(\frac{|M-\kappa S|^2}{\kappa\mu^2}\right) \nonumber \\
 &\simeq&
  -\frac{N_i \kappa^2 \mu^4}{16\pi^2}
  \left[\frac{\kappa}{M}(S+S^\dagger)+\frac{\kappa^2}{2M^2}(S^2+S^{\dagger 2})
  \right] ,
\end{eqnarray}
where
\begin{eqnarray}
 F(x) =
  \frac{N_i}{4} \left[(1+x)^2\log(1+x)^2+(1-x)^2\log(1-x)^2-2x^2\log{x^2}\right].
\end{eqnarray}
Here, $N_i$ is the number of messenger fields that propagate in the
loop diagrams, and $N_i=5$ in our case.  In the second line, the CW
potential is expanded by $S$ around the origin only with the
leading-order contribution about $\kappa\mu^2/M^2$, and constant terms
are neglected. The field value of $S$ on the local minimum is shifted
toward the positive direction, and $m_S^2$ and $m_{f_{\pm}}^2$ become
smaller. Thus, the CW potential makes the local minimum more unstable.
Nevertheless, as long as $\Lambda$ is small enough to satisfy
\begin{eqnarray} \label{eq:constraint_S}
 M \gsim \frac{\kappa^2}{4\pi}\Lambda,
\end{eqnarray}
the local minimum remains and exists near the origin: $\langle S
\rangle \simeq 5 (\kappa^2\Lambda/4\pi M)^2 M/\kappa$.  At this false
vacuum, the condition (\ref{eq:constraint_f}) does not significantly
change, of course, as far as $\langle S \rangle$ is small enough.

Now we discuss the relation between the gravitino mass and the
stability of the metastable SUSY breaking vacuum. First we give the
gravitino mass, normalized by the squark mass, in this model.  The SUSY
breaking is mediated to the visible sector by messenger fields.  The
gaugino and sfermion masses are induced at the messenger scale $\mu_M$
as
\begin{eqnarray} \label{eq:gaugino_mass}
 M_{\tilde g_i}(\mu_M)
  &\simeq& N \frac{\alpha_i(\mu_M)}{4\pi}\frac{\kappa F_S}{M_{\rm mess}}
  \simeq N \frac{\alpha_i(\mu_M)}{4\pi}\frac{\kappa \mu^2}{M}, \\
   M_{\tilde f}^2(\mu_M)
  &\simeq& \sum_i 2 N C_{i} \left( \frac{\alpha_i(\mu_M)}{4\pi} \right)^2
 \left( \frac{\kappa F_S}{M_{\rm mess}} \right)^2
  \simeq \sum_i 2 N C_{i} \left( \frac{\alpha_i(\mu_M)}{4\pi} \right)^2
 \left( \frac{\kappa \mu^2}{M} \right)^2, \label{eq:squark_mass}
\end{eqnarray}
where $C_i$ is the quadratic Casimir, $N$ is the number of the
messenger pair, and $N=1$ for our minimal case.  
We assumed $M \gg \kappa \langle S \rangle$
and $F_S\simeq \mu^2$ in the last form.  Then, the gravitino mass is
given by
\begin{eqnarray} \label{eq:gravitino}
 m_{3/2} \simeq \frac{F_S}{\sqrt{3}M_{\rm pl}}
 \simeq 10~{\rm eV} \left(\frac{M_{\tilde q}}{2~\rm TeV}\right)^2
                   \left(\frac{M}{\kappa\mu}\right)^2 ,
\end{eqnarray}
where $M_{\rm pl} = 2.4 \times 10^{18} $~GeV is the reduced Planck
mass.  Here the mass scale is normalized by the squark mass $M_{\tilde q}$ calculated
by Eq.~(\ref{eq:squark_mass}).

Small $M/\kappa \mu$ is required in Eq.~(\ref{eq:gravitino}) for
the ultra light gravitino. However, the small value of $M/
\kappa \mu$ may lead to the tachyonic messengers, and the large value
of $\kappa$ also destabilizes the local minimum once radiative
corrections are included. The constraints (\ref{eq:constraint_f}) and
(\ref{eq:constraint_S}) lead to the lowerbound on the value of
$M/\kappa\mu$ as
\begin{eqnarray}
 \left( \frac{M}{\kappa\mu} \right)^3 
 \gsim  \frac{1}{4 \pi} \frac{\Lambda}{\mu} .
 \label{Eq:minofm3/2}
\end{eqnarray}
Thus, the ultra light gravitino $m_{3/2} \lesssim 16$~eV is 
marginal.

In this estimate we omitted all possible ${\cal O}(1)$ coefficients.
The precise value of the minimum of $M/\kappa \mu$ is essential to
discuss the possibility of the ultra light gravitino in this model.
In next section we evaluate the decay rate of the false vacuum
numerically and constrain the parameter space of this model.

\section{Stability of metastable vacuum}

As explained in the previous section, the stability of the SUSY
breaking false vacuum is weaker for smaller gravitino mass.  Hence, we
should carefully estimate the transition rate of the false vacuum to
the true ones.  It can be evaluated in the semiclassical techniques
\cite{Coleman}.  In this method, the transition rate of false vacua
described by scalar fields $\phi_i$ is calculated by the Euclidean
path integral $\int [d\phi_i]~e^{-S_E[\phi_i]}$, which is dominated by
the so-called ``bounce'' configuration $\phi_i=\bar{\phi}_i(x)$.  This
bounce is a stationary point of the Euclidean action $S_E[\phi_i]$,
that is $\delta S_E[\bar{\phi}_i]=0$, and it is known to be an $O(4)$
symmetric solution \cite{Coleman:1977th}.  Once we find the bounce
solution, the transition rate per unit volume is estimated as
\begin{eqnarray}
 \Gamma / V  &\simeq& A ~e^{-B}.
\end{eqnarray}
The prefactor $A$ has the fourth power of the typical scale in the potential and
\begin{eqnarray}
 B = S_E[\bar{\phi}_i(r)]-S_E[\phi^f_i],
\end{eqnarray}
where $\phi^f_i$ are constant configurations that continue to stay at the false vacuum.

Several complex scalar fields exist in the present model.  Hence, we
should estimate $S_E$ by finding the bounce solution in a numerical
method.  In the following, we explain our calculation.  For
simplicity, we take all fields real and $f=\bar{f}$. The $D$ term
potential for messenger fields makes this configuration
$|f|=|\bar{f}|$, and the potential for the direction along $f=\bar{f}$
becomes unstable when $S$ is increased.  Then, defining $\phi_1\equiv
f=\bar{f}$ and $\phi_2\equiv S$, we have the following $O(4)$
symmetric Euclidean action,
\begin{eqnarray} 
\label{eq:action}
 S_E[\phi_i(r)]
 = 
  2\pi^2 \int_0^\infty dr~r^3 
  \left[ \sum_{i=1}^2 \frac{k_i}{2} \left(\frac{d \phi_i}{d r}\right)^2
  + V(\phi_i) \right] ,
\end{eqnarray}
where $k_1=4$, $k_2=2$ and $V(\phi_i)$ is the scalar potential, including
the CW potential, that is the sum of
Eqs.~(\ref{eq:potential_tree}) and (\ref{eq:potential_CW}). Then the
bounce solution $\bar{\phi}_i$ is given by solving the following
equations of motion,
\begin{eqnarray} \label{eq:EOM}
 k_i \frac{d^2 \phi_i}{dr^2}+k_i \frac{3}{r}\frac{d \phi_i}{dr}
  = \frac{\partial V}{\partial \phi_i} (\phi_i),
\end{eqnarray}
with boundary conditions,
\begin{eqnarray} \label{eq:BC}
 \lim_{r\rightarrow \infty} \bar{\phi}_i(r) = \phi^f_i,\quad
 \frac{d}{d r} \bar{\phi}_i(r) = 0 .
\end{eqnarray}
Since the prefactor $A$ is approximately given as $A \sim \mu^4$,
$B\gtrsim 400$ is required to make the lifetime of the false vacuum
much longer than the age of the universe. 

To solve the above equations and find the bounce configuration numerically,
we used the calculation method discussed in Ref.~\cite{Konstandin:2006nd}.
We discretized Eqs.~(\ref{eq:EOM}) and (\ref{eq:BC}) as
\begin{eqnarray} 
\label{eq:discretized_EOM}
 &&
 k_i \frac{\phi_{i,j+1}-2\phi_{i,j}+\phi_{i,j-1}}{(\Delta r)^2}
  + k_i \frac{\alpha - 1}{(j+\Delta j)\Delta r}
    \frac{\phi_{i,j+1}-\phi_{i,j-1}}{2 \Delta r}
  = \frac{\partial V}{\partial \phi_i}(\phi_{i,j}),\\
 &&
 \phi_{i,n} = \phi^f_{i},\quad \phi_{i,0}=\phi_{i,1} , 
\end{eqnarray}
where $\phi_{i,j}$ $(j=0,\cdots,n)$ represents the field value of
$\phi_i$ at the $j$-th lattice site and $\Delta r$ is the lattice
size.  These numerical parameters are taken to be $n=200$ and $\Delta
r = 0.25/\mu$ in our calculation.  Parameters $\alpha$ and $\Delta j$
in Eq.~(\ref{eq:discretized_EOM}) are used to find the bounce
configuration in this method.  At first we found the undamped solution
with $\alpha=1$, which can be easily obtained by an improved potential
\cite{Konstandin:2006nd}.  Then we gradually increase $\alpha$ to 4
and decrease $\Delta j$ to 0 with solving the above equations, and
finally the desired bounce solution is obtained\footnote{
In addition, after one bounce solution $\phi^*$ is obtained for a
parameter set, new bounce solutions for the new
parameter sets are also iteratively derived from it. 
In this derivation the right-hand
side of Eq.~(\ref{eq:discretized_EOM}) is expanded as
\begin{eqnarray}
 \frac{\partial V}{\partial \phi_i}(\phi_{i,j})
 \simeq
  \frac{\partial V}{\partial \phi_i}(\phi^*_{i,j})
  +
  \sum_{k=1}^2 (\phi_{k,j}-\phi^*_{k,j}) 
  \frac{\partial^2 V}{\partial \phi_i\partial \phi_k}(\phi^*_{i,j}) ,
\end{eqnarray}
to linearized the system of equations.
}.
From this bounce configuration, the coefficient $B$ is calculated by
the following discretized Euclidean action,
\begin{eqnarray}
 S_E[\phi_{i,j}]
 = 
  2\pi^2 \sum_{j=0}^{n-1} \left(j+\frac{1}{2}\right)^3 \Delta r^3
  \left[ 
   \sum_{i=1}^2 \frac{k_i}{2} 
   \left(\frac{\phi_{i,j+1}-\phi_{i,j}}{\Delta r}\right)^2
   + \frac{V(\phi_{i,j+1}) + V(\phi_{i,j})}{2}
  \right].
\end{eqnarray}

\begin{figure}[t]
\begin{center}
\begin{tabular}{cc}
\includegraphics[scale=0.6, angle = 0]{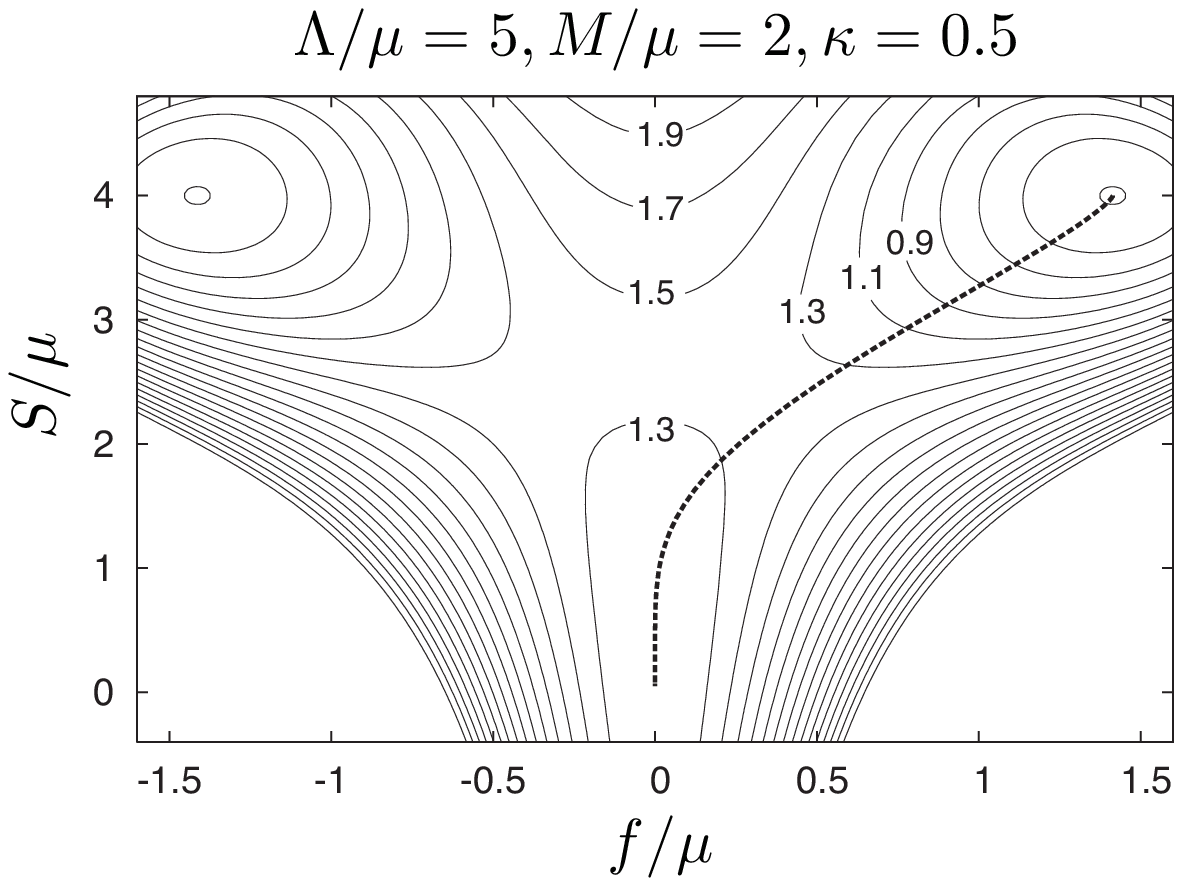}&
\includegraphics[scale=0.6, angle = 0]{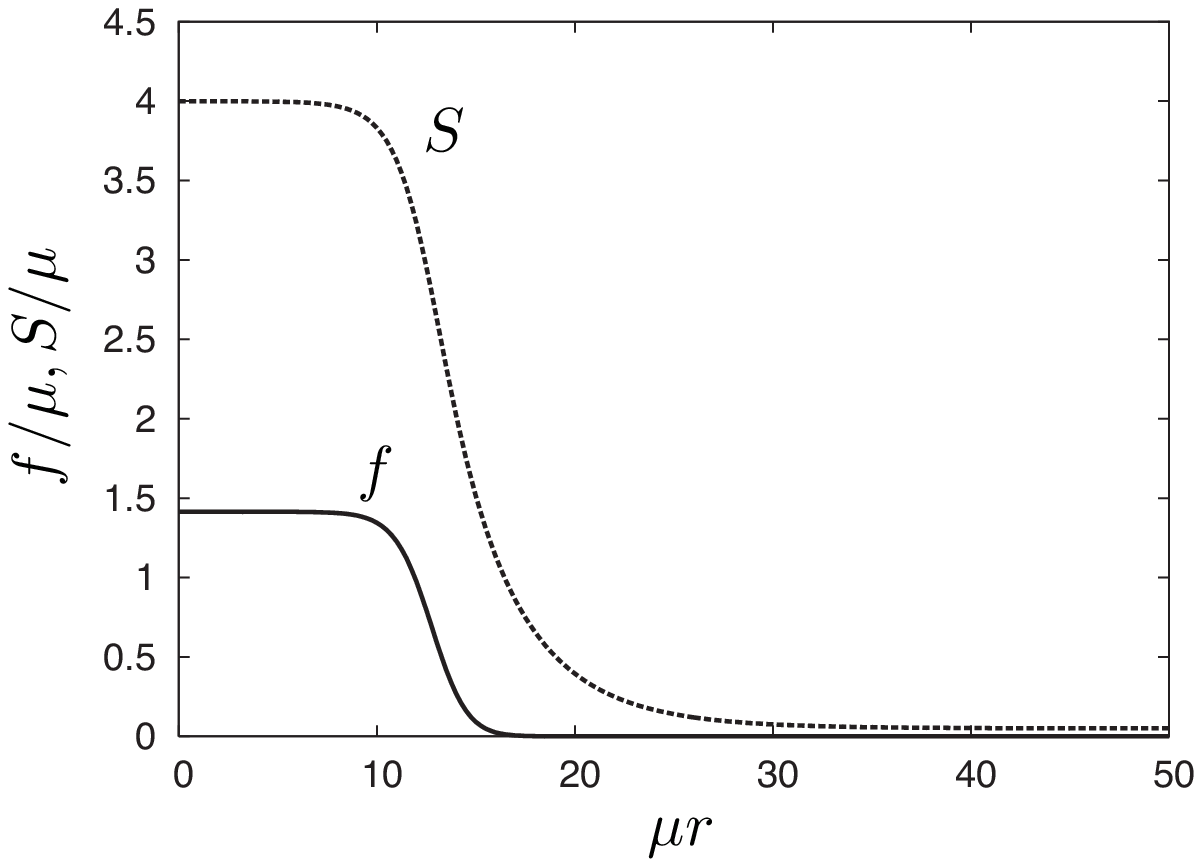}
\end{tabular}

\caption{\label{fig:bounce} 
A typical shape of potential
  $V(\phi_i)$ (left) and dependence of bounce solution $\bar{\phi}_i(r)$ on $r$
  (right).  Parameters are taken as $\Lambda/\mu=5$, $M/\mu=2$ and
  $\kappa=0.5$.  A trajectory of $\bar{\phi}_i(r)$ is shown
  by a dotted line in the left figure. Height of the potential 
  in the figure is scaled by $\mu^4$.   
}
\end{center}
\end{figure}

In Fig.~\ref{fig:bounce}, we show a typical potential shape (left) and
the corresponding bounce configuration derived by the above procedure
(right) for $\Lambda/\mu=5$, $M/\mu=2$ and $\kappa=0.5$.  In the left
figure, a trajectory between the false vacuum and true one is
shown by a dotted line.  In the right figure, the trajectories of
$\phi_1\,(=f)$ and $\phi_2\,(=S)$ are shown by solid and dotted lines,
respectively.  They are rather simple in most of the parameter space
as long as the conditions Eqs.~(\ref{eq:constraint_Lambda}),
(\ref{eq:constraint_f}) and (\ref{eq:constraint_S}) are satisfied.
First $\phi_2$ moves toward the positive direction and
$\phi_1$ is almost unchanged.  Then around $\phi_2\simeq
(M-\sqrt{\kappa}\mu)/\kappa$, the messenger field becomes tachyonic
and both fields start to go toward one of the true vacua.

\begin{figure}[t]
\begin{center}
\begin{tabular}{cc}
\includegraphics[scale=0.6, angle = 0]{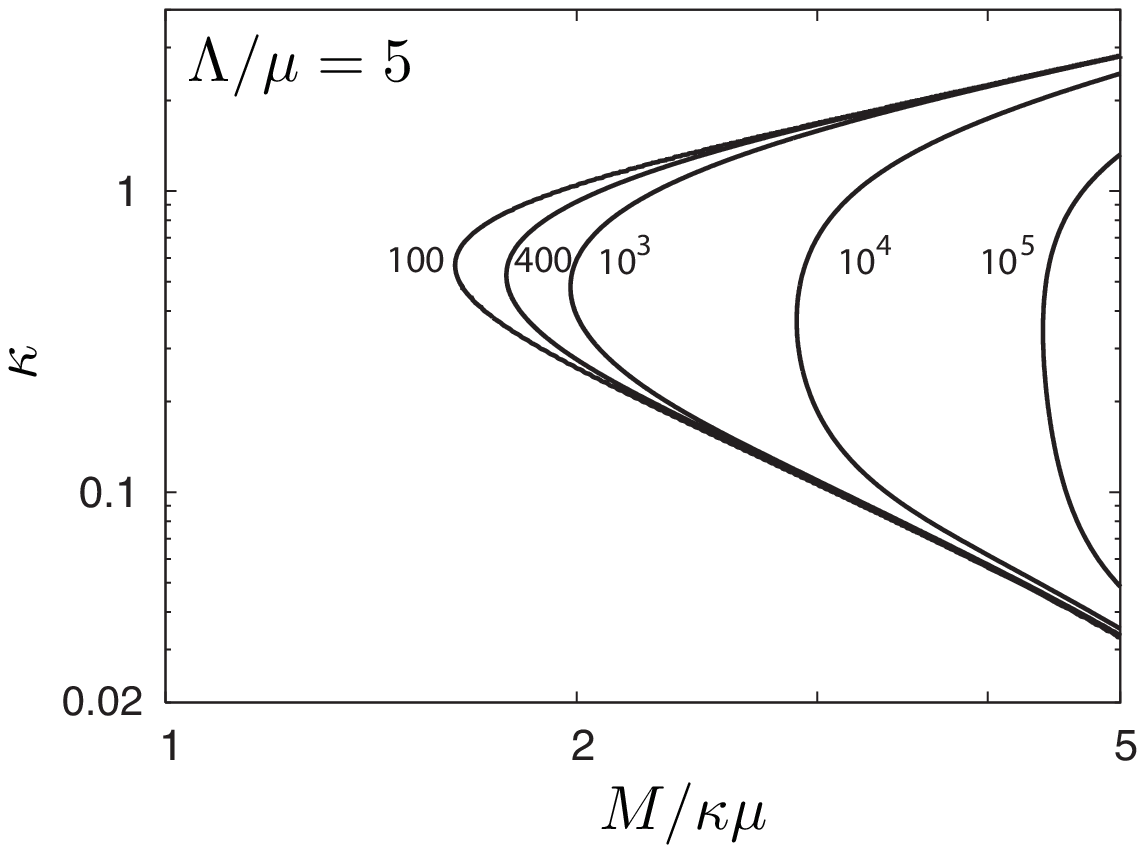}&
\includegraphics[scale=0.6, angle = 0]{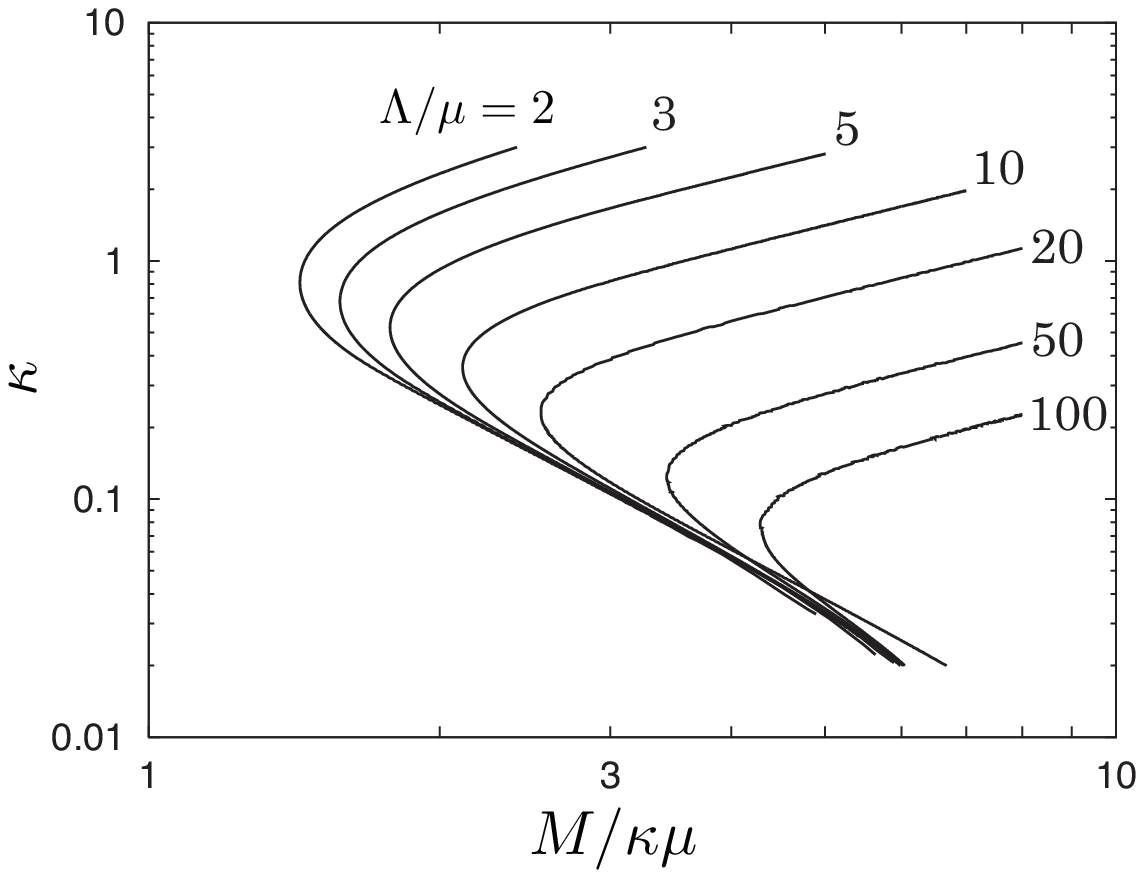} 
\end{tabular}
\caption{\label{fig:contour} 
Values of $B$ for $\Lambda/\mu=5$ (left)
  are shown dependent on $ M / \kappa \mu $ and $\kappa$.  Constraints
  on the parameter spaces by $B\ge400$ for several values of
  $\Lambda/\mu$ (right).
}
\end{center}
\end{figure}

Next, we discuss the parameter dependence of the decay rate of the
false vacuum.  The potential $V(\phi_i)$ are described by 4
parameters: $\Lambda$, $M$, $\mu$, and $\kappa$.  Since the overall
mass scale is irrelevant to the calculation of $B$, we normalized all
dimensional parameters by $\mu$.  In Fig.~\ref{fig:contour}~(left), we
show the value of $B$ as a function of $M/\kappa\mu$ and $\kappa$ for
$\Lambda/\mu=5$.  In the lower-left region the vacuum is unstable
along the $f (\bar f)$ direction since the condition, $M^2 > \kappa
\mu^2$, in Eq.~(\ref{eq:constraint_f}) is not satisfied. In the
upper-left region the CW potential is not negligible and the vacuum is
unstable due to the violation of the condition in
Eq.~(\ref{eq:constraint_S}). From this figure, it is found that $B$ is
determined mainly by the value of $M/\kappa\mu$ except around the
boundaries given in Eqs.~(\ref{eq:constraint_f}) and
(\ref{eq:constraint_S}).

In Fig.~\ref{fig:contour}~(right), the allowed regions by the
constraint $B \ge 400$ are shown for several values of $\Lambda/\mu$.
The allowed regions are the right side of solid lines.  The minimum
allowed value for $M/\kappa\mu$ is almost determined by the
intersecting point of the stability conditions,
Eqs.~(\ref{eq:constraint_f}) and (\ref{eq:constraint_S}) for fixed
$\Lambda/\mu$.  Hence, the bound is roughly given by
Eq.~(\ref{Eq:minofm3/2}).
The difference between Eq.~(\ref{Eq:minofm3/2}) and the bound derived
by Fig.~\ref{fig:contour} is explained by two contributions.  First,
the numerical calculation of the transition rate cuts some region near
this bound as shown in Fig.~\ref{fig:contour} (left).  The other is
the approximation neglecting ${\cal O}$(1) factors in the calculation
of Eq.~(\ref{eq:constraint_S}). The rest condition
(\ref{eq:constraint_Lambda}) gives upperbound on $M/\kappa \mu$ with
a constant $\Lambda / \mu $.  Hence, this condition is irrelevant to
the lowerbound on the gravitino mass.

The parameter dependence of $B$ is understood from an approximate estimate
based on Ref.~\cite{Duncan:1992ai}, where the
vacuum transition rate is calculated for a triangle-shaped potential
of one real scalar field.  Fortunately, the trajectory of the bounce solution $\bar{\phi}_i(r)$ is rather simple as depicted in
Fig.~\ref{fig:bounce}, except around the boundaries,
Eqs.~(\ref{eq:constraint_f}) and (\ref{eq:constraint_S}).  We
approximated the potential along this trajectory by a triangle form
and derived as
\begin{eqnarray} \label{eq:B_triangle}
 B \simeq \frac{8\pi^2 M^4}{\kappa^4 \mu^4} ,
\end{eqnarray}
where this estimate is applicable for thin wall region\footnote{
Our estimate is different from that in Ref.~\cite{SimpGMSB},
since the result in Ref.~\cite{SimpGMSB} can be realized using a
different assumption near the boundaries.}.
This approximated estimate explains some features of numerical
results.  First, the value of $B$ depends mainly on the combination of
$M/\kappa \mu$ and the power of $M/\kappa \mu$ is rather large.
Second, the dependence of $B$ on $\Lambda$ is rather weak.  This
implies that $B$ does not change so much for different values of
$\Lambda$ as long as $M/\kappa\mu$ are fixed, although the allowed
region are changed.

\begin{figure}[t]
\begin{center}
\begin{tabular}{cc}
\includegraphics[scale=0.6]{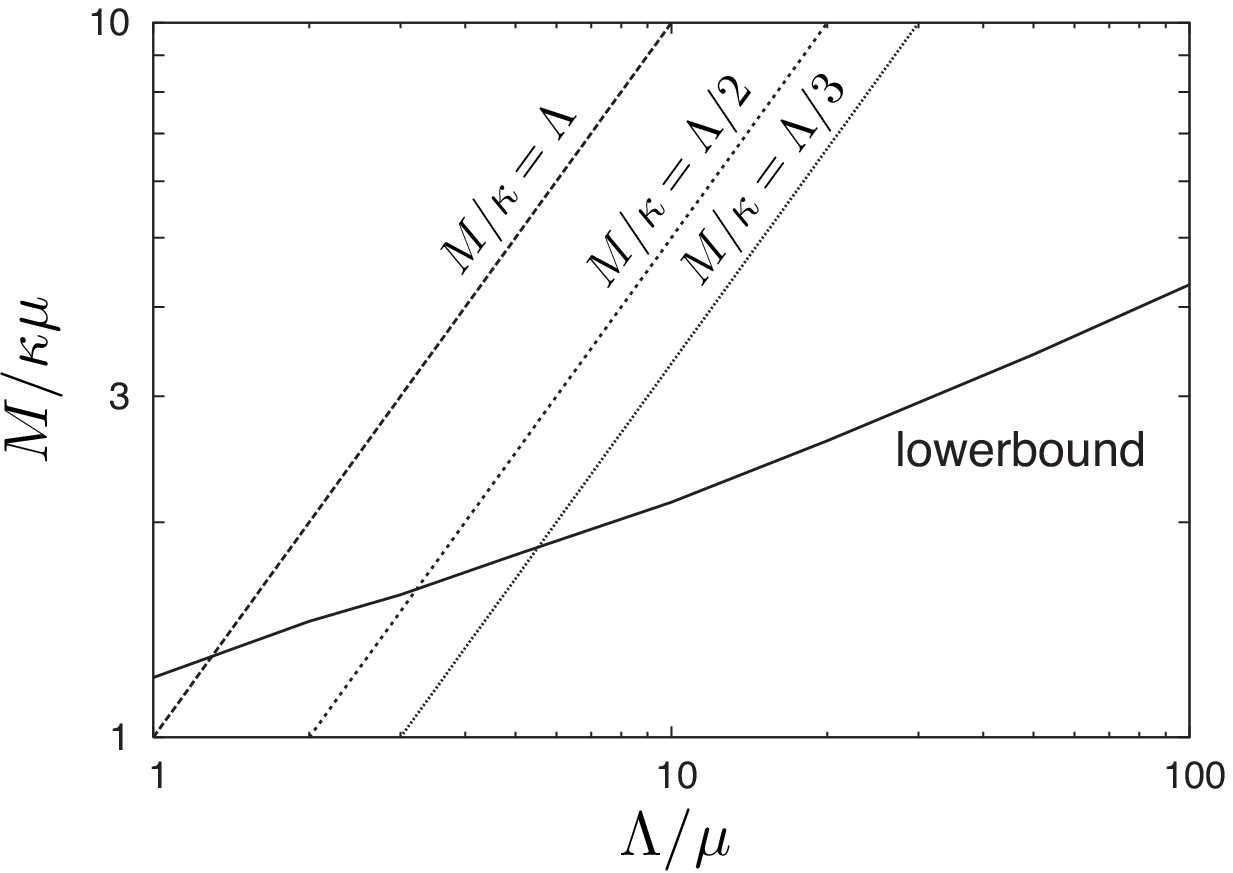}&
\includegraphics[scale=0.6]{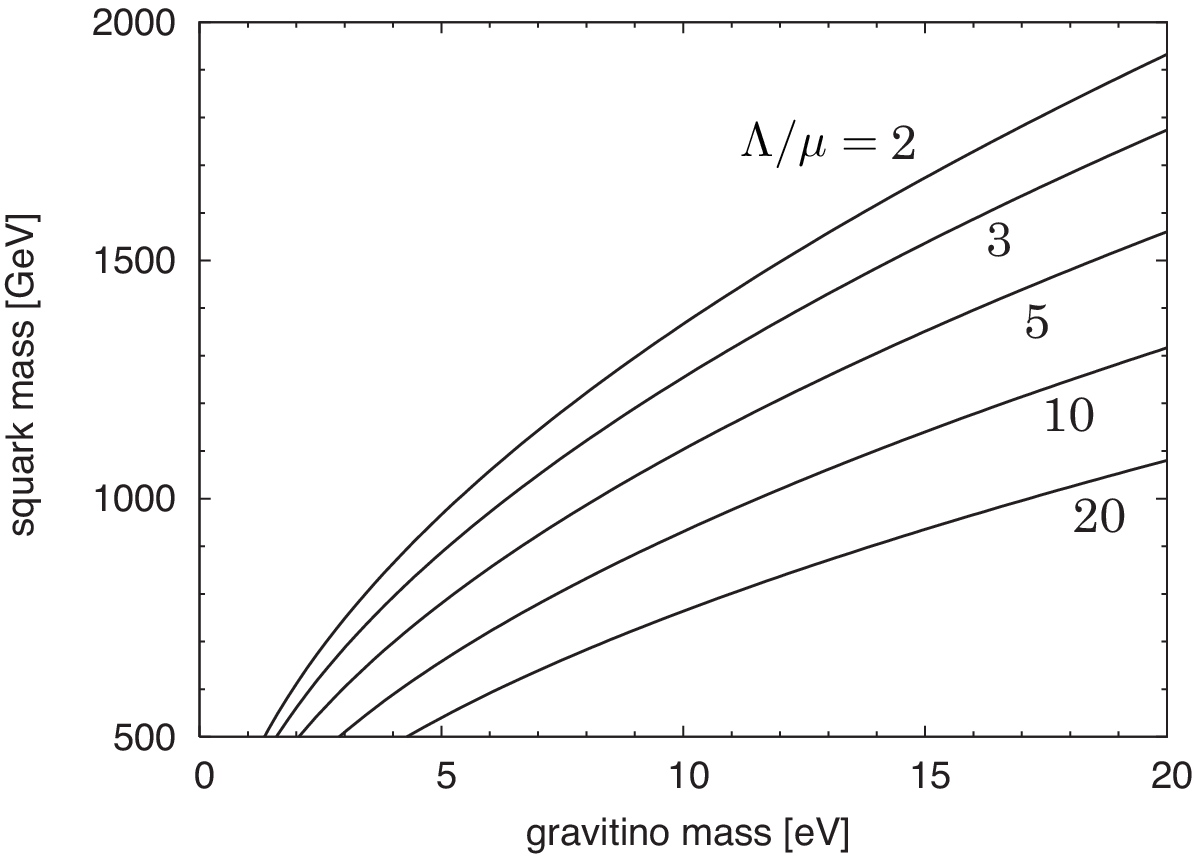} 
\end{tabular}
\caption{\label{fig:constraint} 
Constraints on $M/\kappa\mu$ as 
  functions of $ \Lambda / \mu $ (left).  Solid line is lowerbound on
  $M/\kappa \mu$, while dotted lines show $M/\kappa \Lambda = 1,1/2,1/3$
  from left to right.  Upperbound for the squark mass as a function of
  the gravitino mass (right).  We take $\Lambda / \mu = 2,3,5,10,20$.
  Allowed regions are below the lines.
}
\end{center}
\end{figure}

In Fig.~\ref{fig:constraint}~(left), we summarized the allowed minimum
values of $M/\kappa \mu$ dependent on $\Lambda/\mu$.  The solid line
shows the lowerbound on $M/\kappa \mu$ and the dotted lines are
$M/\kappa \Lambda=1, 1/2$ and $1/3$, respectively.  For larger values of
$M/\kappa \Lambda$, the description as an effective low-energy theory
becomes inadequate and higher order terms neglected in
Eq.~(\ref{eq:kahler}) become important.  In these cases, UV models
are needed for complete analysis.

The lowerbound on $M/\kappa\mu$ can be translated to the constraints
on the gravitino and squark mass through Eq.~(\ref{eq:gravitino}).
In Fig.~\ref{fig:constraint}~(right), we show the allowed region for
the squark and gravitino masses for several values of $\Lambda/\mu$,
where the allowed regions are below lines.  From this figure, the
upperbound on the squark mass for fixed value of $ m_{3/2}$ is
obtained.  It is found that the relatively light squark mass $M_{\tilde q}
\lesssim 1800$~GeV are required to have $m_{3/2}\lesssim
16~{\rm eV}$ even if $\Lambda/\mu=2$. When $\Lambda$ is larger, the 
constraint is more severe.
 The ultra light gravitino in this model may be checked,
since the mass range is accessible in the LHC experiments.
This result is derived by using the approximation of the last form
of Eq.~(\ref{eq:gravitino}).
The upperbound on the squark mass is slightly relaxed without this approximation.
Nevertheless, we checked that the difference is about 10\,-\,20~\%.

The LEP experiments constrain the SM-like Higgs boson mass as $m_h >
114.4$~GeV \cite{PDG}.  For the large enough Higgs boson mass, the stop mass
should be larger than about 1000~GeV to have a large radiative
correction through top and stop loop diagrams.  This suggests that the
cutoff scale is low enough to satisfy $\Lambda/\mu \lesssim 20$  as
in Fig.~\ref{fig:constraint}.  This is reduced to the bound,
$\Lambda \lesssim 10^{7}~{\rm GeV}$,
with $\mu \simeq 200 ~{\rm TeV}( m_{3/2} / 10~\rm eV)^{1/2}$.
This gives a strong implication for UV models.

Finally we briefly discuss the case of non-minimal messenger fields
$(N>1)$, which is a quite interesting situation, since it may be
possible to measure the gravitino mass in future collider experiments
\cite{Hamaguchi:2007ge}.  If $N$ copies of ${\bf 5} + {\bf 5}^*$ are
introduced, the gluino and squark masses become larger and scale like
$N$ and $\sqrt{N}$, respectively.  On the other hand, the SUSY
breaking vacuum becomes more unstable, since the CW
potential in Eq.~(\ref{eq:potential_CW}) is proportional to $N$.  A
factor $N^{1/6}$ appears in the lowerbound on $M/\kappa\mu$ for a
fixed $\Lambda/\mu$.  Then the upperbounds on the gluino and squark
masses scale like $N^{5/6}$ and $N^{1/3}$ respectively.  Hence, squark
may be still detected at the LHC experiments even for $N>1$.

\section{Summary }

In this paper we have discussed the possibility that the ultra light
gravitino with $m_{3/2} \lesssim 16~{\rm eV}$ is realized in the GMSB
model with the metastable SUSY breaking vacuum.  In our numerical
calculation of the decay rate of the false vacuum, the
Coleman-Weinberg potential is included in the scalar potential, since
it destabilizes the metastable SUSY breaking vacuum.  We have found
that when this model predicts a rather light gravitino, an upperbound
on the squark mass $M_{\tilde q} \lesssim 1800~{\rm GeV}$ 
 at most is obtained for the minimal messenger fields from the
stability of the vacuum. They are detectable in the LHC experiments,
and hence this scenario may be checked.  Even for the cases of $N>1$,
squarks are still relatively light.  We also found that to have the
ultra light gravitino the cutoff scale must be smaller than $
10^{7}$~GeV, which is an important requirement for UV
models that realize this scenario.

\section*{Acknowledgments}

The works of JH and MS are supported in part by the Grant-in-Aid
for Science Research, Ministry of Education, Science and Culture,
Japan (No.~19034001 and No.~18034002 for JH and No.~18840011 for MS).
Also, that of MN  is supported in part by JSPS.

\end{document}